\documentclass{elsart}
\usepackage{epsfig}

\def\beq{\begin{equation}} \def\eeq{\end{equation}}
\def\beqa{\begin{eqnarray}} \def\eeqa{\end{eqnarray}}
\def\bce{\begin{center}}  \def\ece{\end{center}}
\def\bfig{\begin{figure}}  \def\efig{\end{figure}}
\def\bit{\begin{itemize}}    \def\eit{\end{itemize}}
\def\ben{\begin{enumerate}}    \def\een{\end{enumerate}}
\def\xs{cross-section} \def\xss{cross-sections}
\def\npb{Nucl. Phys. {\bf B}} \def\plb{Phys. Lett. {\bf B}}
\def\prd{Phys. Rev. {\bf D}}
\def\prp{Phys. Rep.} 
\def\cl{{\mathcal L}}

\begin{document}

\newlength{\caheight}
\setlength{\caheight}{12pt}
\multiply\caheight by 7
\newlength{\secondpar}
\setlength{\secondpar}{\hsize}
\divide\secondpar by 3
\newlength{\firstpar}
\setlength{\firstpar}{\secondpar}
\multiply\firstpar by 2

\hfill
\parbox[0pt][\caheight][t]{\secondpar}{
  \rightline
  {\tt \shortstack[l]{
    FNT/T-00/01
  }}
}

\begin{frontmatter}

\title{Anomalous quartic couplings in six-fermion\\ processes at the
       Linear Collider}

\author{Fabrizio Gangemi}

\address{Dipartimento di Fisica Nucleare e Teorica, Universit\`a di Pavia}
\address{and Istituto Nazionale di Fisica Nucleare, Sezione di Pavia}
\address{Via A.~Bassi 6, I--27100 Pavia, Italy}

\begin{abstract}
The dimension-four genuine anomalous quartic couplings are studied in
processes of six-fermion production via $e^+e^-$ collisions. Complete
tree-level electroweak calculations are performed including
initial-state-radiation and beamstrahlung. The analysis of final-state
distributions can be used to find kinematical cuts to enhance the effects of
anomalous couplings. For the parameters of the custodial-symmetry-conserving
anomalous couplings a sensitivity in the range between $10^{-3}$ and $10^{-2}$
can be expected at 1 TeV.
\end{abstract}


\end{frontmatter}

\section{Introduction}
\label{sect:intro}
The experimental measurement of the gauge self-couplings is an important test
of the Standard Model (SM), that is still at the beginning at present colliders.
The trilinear gauge couplings have recently become a subject of studies at
LEP II and Tevatron (for recent results, see \cite{tamp99}), and will also be
measured at the Linear Collider (LC) \cite{lc}. The quartic couplings are very
weakly constrained by present experimental data, through loop
diagrams~\cite{agcloop}, while the observation of quartic gauge coupling effects
at tree-level at LEP is rather difficult when photons are involved~\cite{qgcph}
and is completely outside the reach of LEP when only massive gauge bosons are
present.

The determination of quadrilinear gauge couplings is of particular interest in
connection with the problem of the electroweak symmetry breaking. Indeed, if a
Higgs boson lighter than 1 TeV does not exist, the tree-level amplitudes for
the gauge boson scattering in the Standard Model are known to violate the
unitarity limits at tree-level, indicating the existence of strong interactions.
This scenario can be the result of different models of symmetry breaking and may
be described in the general framework of effective lagrangians, where the
electroweak interactions in the low-energy limit are represented by one
general parameterization \cite{chlbook}. Different values of the parameters
correspond to different models. The effective lagrangian is organized as a power
series in $p/4\pi v$, where $p$ is the scale of momenta of the phenomena under
study and $v\simeq 246$ GeV is the scale of symmetry breaking. The lowest-order
terms of this expansion are model-independent. The higher order terms involve
new gauge coupling structures that give rise to anomalous vertices and contain
coefficients that are model-dependent.

It is worth noticing that the lowest-order non-trivial quartic couplings in the
effective lagrangian (that are the $O(p^4)$ or dimension-four terms) involve
only massive gauge bosons and thus their effects can be observed either in
loop contributions, or at tree-level in processes with at least six fermions
in the final state.

The measurement of the parameters describing the gauge self-couplings, that
gives the opportunity of constraining some possible models of new physics,
will be an important objective of the LC, where significant
improvements will be possible with respect to present colliders for various
reasons. In the first place, the c.m. energy is sufficiently high for the
production of up to three gauge bosons in the final state, making it possible
to study the quadrilinear couplings at tree-level. Moreover, since the
deviations from the SM values of these couplings destroy the unitarity
cancellations, their effect is enhanced at high energy. Finally, the special
features of the LC, such as the high luminosity and the possibility of having
polarized $e^+$ and $e^-$ beams will provide a very high sensitivity to the
anomalous couplings.

Several phenomenological studies have been performed on the possibility of
constraining the anomalous quartic couplings at $e^+e^-$ colliders in the
approximation of real vector bosons in the final state, both for the couplings
including photons~\cite{qgcph} and for those involving gauge bosons
only~\cite{qgcm,qgcm1}. In the latter case, however, for more realistic
predictions, as observed above, processes with at least six fermions in the
final state have to be examined. This kind of processes also allows for the
analysis of final-state distributions that are not accessible in the real
approximation, and that can give useful information.

The objective of the present study is to analyse a class of six-fermion ($6f$)
processes where the genuine quartic gauge couplings of dimension four are
involved. The results that will be shown have been obtained by means of
complete tree-level calculations in the framework of the chiral approach to
electroweak interactions.

In Section~\ref{sect:2} the theoretical framework and some technical details
of the calculation are explained. In Section~\ref{sect:3} the numerical results
are presented and discussed, and Section~\ref{sect:4} contains the conclusions.

\section{Theoretical framework and calculation technique}
\label{sect:2}
The construction of the effective lagrangian for the electroweak interactions
satisfying the requirement of $SU(2)\times U(1)$ gauge invariance may be
found in the literature~\cite{ewchl}. It will be useful to mention here that
in this approach the lagrangian can be written in the form
\beq
\label{eq:2.0}
\cl=\cl_{YM}+\cl_F+\cl_S\ ,
\eeq
where $\cl_{YM}$ is the Yang-Mills lagrangian,
\beq
\label{eq:2.01}
\cl_{YM}=-\frac{1}{4}W^i_{\mu\nu}W^{i\mu\nu}-\frac{1}{4}B_{\mu\nu}B^{\mu\nu}\ ,
\eeq
$\cl_F$ is the fermionic contribution and $\cl_S$ is the scalar contribution.
The standard couplings between gauge bosons are given by the Yang-Mills
lagrangian. The scalar sector is represented by means of the unitary matrix
\beq
\label{eq:2.1}
U=\exp\left({i\frac{{\tau_i}{\pi_i}}{v}}\right)\ ,
\eeq
where $v=246$ GeV and $\pi_i$ are the would-be Goldstone bosons.
The gauge-invariant operators contributing to $\cl_S$ can be constructed by
taking the traces of the following building blocks:
\beqa
\label{eq:2.2}
T&=&U\tau_3U^\dagger\\
\label{eq:2.3}
V_\mu&=&(D_\mu U)U^\dagger\\
\label{eq:2.4}
W_{\mu\nu}&=&\partial_\mu W_\nu-\partial_\nu W_\mu+
ig[W_\mu,W_\nu]\ ,
\eeqa
where
$D_\mu U=\partial_\mu U+\frac{ig}{2}\tau_iW^i_\mu U-\frac{ig'}{2}B_\mu U\tau_3$.
At leading order, assuming the custodial symmetry, the following contribution
to $\cl_S$ is found:
\beq
\label{eq:2.5}
\cl_0=\frac{v^2}{4}Tr\left((D_\mu U)^\dagger D^\mu U\right)\ ,
\eeq
that gives the masses to the $W^\pm$ and $Z$ bosons.

At next-to-leading order, that corresponds to dimension
four, new couplings appear, and the independent gauge-invariant structures are
multiplied by coefficients that are model-dependent. At this order, trilinear
and quadrilinear vertices are present. The operators that give rise to
quadrilinear and not to trilinear vertices, with the condition of $CP$
conservation, are (using the standard notation):
\beqa
\nonumber
\cl_4&=&\alpha_4(Tr(V_\mu V_\nu))^2\\
\nonumber
\cl_5&=&\alpha_5(Tr(V_\mu V^\mu))^2\\
\nonumber
\cl_6&=&\alpha_6Tr(V_\mu V_\nu)Tr(TV^\mu)Tr(TV^\nu)\\
\nonumber
\cl_7&=&\alpha_7Tr(V_\mu V^\mu)(Tr(TV_\nu))^2\\
\label{eq:2.5a}
\cl_{10}&=&\alpha_{10}\frac{1}{2}(Tr(TV_\mu)Tr(TV_\nu))^2\ .
\eeqa
The custodial symmetry is respected only by the operators $\cl_4$ and $\cl_5$,
and is violated by the other three operators, due to the presence of $T$.
In the unitary gauge the above terms are given by:
\beqa
\nonumber
\cl_4&=&
\alpha_4g^4\left(\frac{1}{2}W^+_\mu W^{+\mu}W^-_\nu W^{-\nu}
+\frac{1}{2}(W^+_\mu W^{-\mu})^2\right.\\
\label{eq:2.6}
&+&\left.\frac{1}{c_W^2}W^+_\mu Z^\mu W^-_\nu Z^\nu+
\frac{1}{4c_W^4}(Z_\mu Z^\mu)^2\right)\\
\nonumber
\cl_5&=&
\alpha_5\left((W^+_\mu W^{-\mu})^2+\frac{1}{c_W^2}W^+_\mu W^{-\mu} Z_\nu Z^\nu
\right.\\
\label{eq:2.7}
&+&\left.\frac{1}{4c_W^4}(Z_\mu Z^\mu)^2\right)\\
\label{eq:2.8}
\cl_6&=&
\alpha_6g^4\left(\frac{1}{c_W^2}W^+_\mu Z^\mu W^-_\nu Z^\nu+
\frac{1}{2c_W^4}(Z_\mu Z^\mu)^2 \right)\\
\label{eq:2.9}
\cl_7&=&
\alpha_7g^4\left(\frac{1}{c_W^2}W^+_\mu W^{-\mu} Z_\nu Z^\nu+
\frac{1}{2c_W^4}(Z_\mu Z^\mu)^2\right)\\
\label{eq:2.10}
\cl_{10}&=&
\alpha_{10}\frac{g^4}{2c_W^4}(Z_\mu Z^\mu)^2\ .
\eeqa
As can be seen, these operators give the anomalous contributions to the
quadrilinear gauge couplings involving the $W^\pm$ and $Z$ vector bosons, to be
added to the standard ones that are contained in the Yang-Mills lagrangian. In
particular, it is easy to find, in the above formulas, anomalous contributions
to the vertices $4W$ ($\cl_4$ and $\cl_5$), $WWZZ$ ($\cl_4$, $\cl_5$, $\cl_6$
and $\cl_7$) and $4Z$ (all the operators).

In this paper, the numerical results of a phenomenological study on $6f$
processes involving these anomalous couplings will be presented.
These results have been obtained by means of a computer code already employed
in other $6f$ analyses~\cite{6fhiggs,6ftop}. In this code the scattering
amplitudes are calculated by the automatic algorithm ALPHA~\cite{alpha},
and the Monte Carlo integration procedure is the result of an adaptation of the
four-fermion codes HIGGSPV~\cite{higgspv} and WWGENPV~\cite{wwgenpv} to the
$6f$ calculations.
For the present study, the lagrangian in eq.~(\ref{eq:2.0}) with
$\cl_S=\cl_0+\cl_4+\cl_5+\cl_6+\cl_7+\cl_{10}$ has been implemented. The case
in which the coefficients $\alpha_i$ are all equal to zero is considered as a
reference model, that represents the limit of infinite Higgs mass in the SM at
tree-level, and the deviations from such a model when the coefficients are
non-vanishing have been studied.

The input parameters are $G_\mu$, $M_W$ and $M_Z$. The widths of the $W$ and
$Z^0$ bosons and all the couplings are calculated at tree-level. All the
fermions are massless. For the propagators of the gauge bosons, the
``fixed-width'' scheme has been adopted. The reliability of this approach in
$6f$ calculations at energies of the order of the TeV has been discussed in
detail in ref.~\cite{6fhiggs}.

The aim of the present work is to provide an analysis, that has never been done
till now, of the impact of a complete electroweak $6f$ calculation on the study
of anomalous quartic couplings. The numerical results contained in the following
section are a first illustrative application of the new implementation of
anomalous couplings in the code mentioned above, and the processes considered,
although not giving a complete picture of the anomalous quartic coupling
phenomenology, are relevant due to their sizeable \xss, as can be seen below.

\section{Physical processes and results}
\label{sect:3}
The processes $e^+e^-\to 2q+2q'\nu_e\bar\nu_e$, with $q=u,c$ and $q'=d,s$, have
been considered. This choice is motivated by the two objectives of having
contributions from all possible quartic vertices involved in the anomalous
terms under consideration, and of having no more than two neutrinos in the
final state. The first point can be explained by considering the signature
$u\bar ud\bar d\nu_e\bar\nu_e$: as can be seen in fig.~\ref{fig:agcd0}, that
shows the diagrams with a quadrilinear vertex, this signature includes
contributions from the $4W$, $WWZZ$ and $4Z$ vertices.
\bfig
\bce
\epsfig{file=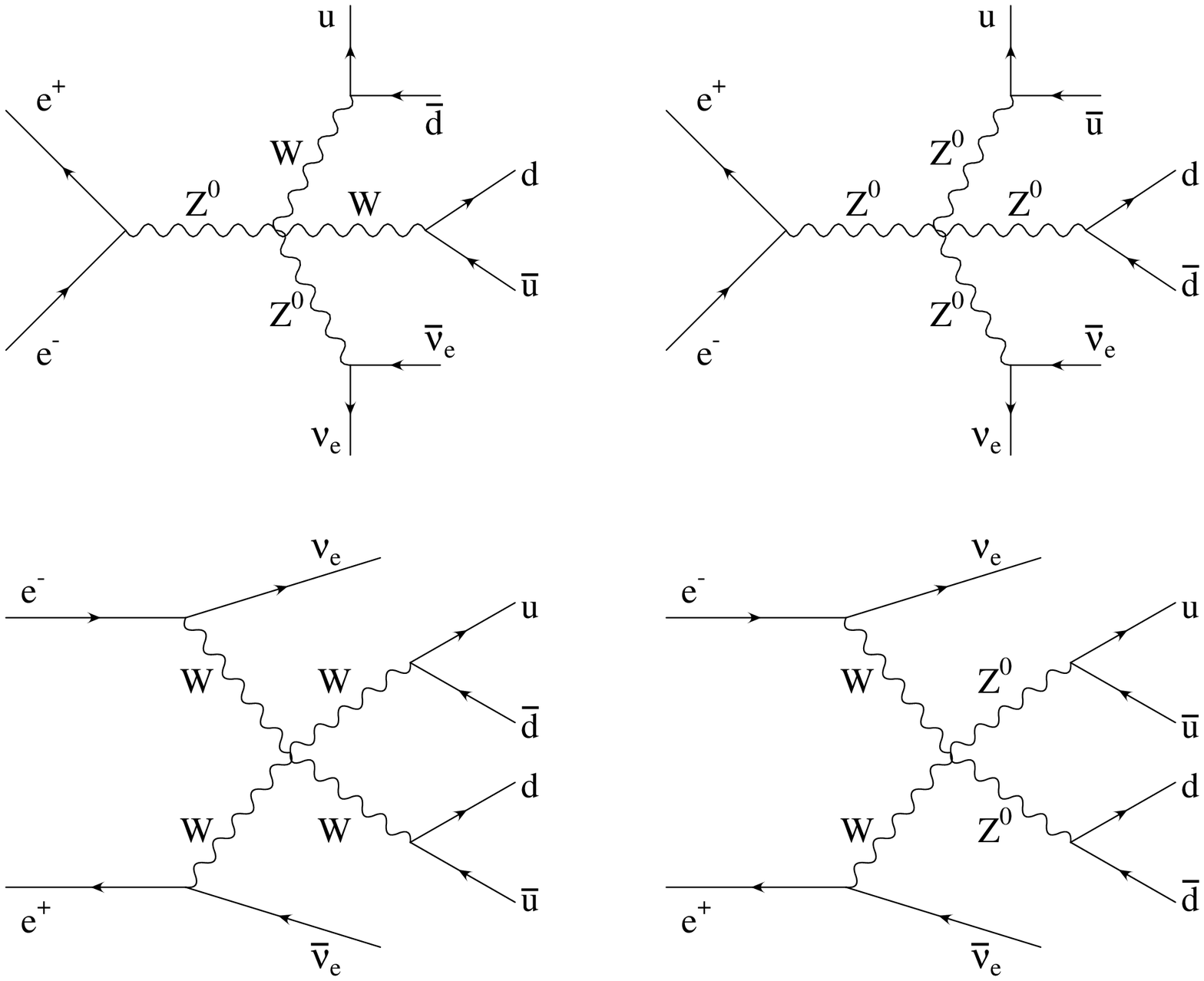,height=7.5cm,width=9.1cm}
\caption{\small Diagrams with quadrilinear massive gauge boson vertices
contributing to the process $e^+e^-\to u\bar ud\bar d\nu_e\bar\nu_e$. Notice
that the diagram on the right in the first row (with a $4Z$ vertex) can be only
anomalous, while the others have both standard and anomalous contributions.}
\label{fig:agcd0}
\ece
\efig
From the point of view of the calculation, in order to obtain the correct sum
over colours, with the version of ALPHA used here, that does not contain the
colour degrees of freedom, it is necessary to apply a procedure analogous to
the one already illustrated in ref.~\cite{6ftop}. This procedure consists of
the combination, with proper weights, of the results from the different
processes and is valid in the approximation of massless quarks and unit CKM
matrix. The signatures $u\bar d\bar cs\nu_e\bar\nu_e$,
$u\bar ud\bar d\nu_e\bar\nu_e$ and $u\bar us\bar s\nu_e\bar\nu_e$ must be
combined according to the following formula:
\beq
\label{eq:3.1}
\sigma_{uudd} + \sigma_{udcs} + \sigma_{uuss}
=N_c(\hat\sigma_{uudd} + (2N_c-1)(\hat\sigma_{udcs}+\hat\sigma_{uuss}))\ ,
\eeq
where the results directly provided by ALPHA are denoted by $\hat\sigma$,
while the \xss\ including the colour degrees of freedom are denoted by
$\sigma$. The interested reader is referred to \cite{6ftop} for more details.

The signatures $c\bar s\bar ud\nu_e\bar\nu_e$, $c\bar cs\bar s\nu_e\bar\nu_e$
and $c\bar cd\bar d\nu_e\bar\nu_e$ are obtained from the previous ones by means
of the simultaneous exchanges $u\leftrightarrow c$ and $d\leftrightarrow s$,
and, given the set of parameters adopted here, they give exactly the same
result, so that they can be easily taken into account by including a factor of
2 in eq.~(\ref{eq:3.1}).

In principle, assuming that the $b$-tagging technique is applied, and thus the
$b$ quark can be identified, a realistic calculation should take into account
all the possible combinations of the remaining flavours, $u$, $d$, $c$ and $s$
that cannot be distinguished. These combinations include, in addition to those
mentioned above, the final states $u\bar uc\bar c\nu_e\bar\nu_e$,
$u\bar uu\bar u\nu_e\bar\nu_e$ $d\bar ds\bar s\nu_e\bar\nu_e$
$d\bar dd\bar d\nu_e\bar\nu_e$ and those obtained from these with the exchanges
$u\leftrightarrow c$ and $d\leftrightarrow s$. Moreover a sum over the neutrino
flavours should be made. Such signatures have been neglected in this first
analysis. Nevertheless, it will be possible to examine the most important
qualitative features of the phenomenology of anomalous gauge couplings in
$6f$ processes, while referring to further developments for a better
accuracy from a quantitative point of view.

In the following, the results of complete electroweak calculations at tree level
are presented, where the above specified class of processes is considered.

As can be seen in the scattering of real gauge bosons, as a consequence of
the absence of the Higgs boson and of the gauge cancellations that ensure
unitarity, the \xss\ at high energy can be expected to increase rapidly and
to violate the unitarity bounds. The energy at which unitarity is violated
depends on the values of the parameters $\alpha_i$: these should then be taken
within some bounds if the unitarity condition has to be respected at a given
energy. However, it is useful to make a first rough analysis of the dependence
of the \xss\ on the anomalous parameters in a wide range, while the sensitivity
to the anomalous couplings in a smaller range will be studied in a second step.
Thus, in the first set of results, the \xs\ has been evaluated at the two c.m.
energies of 500 GeV and 1 TeV, and allowing each of the coefficients $\alpha_i$
to vary in the range $(-0.2,0.2)$, while keeping all the others equal to zero.
A simple set of kinematical constraints has been adopted, by requiring the
invariant masses of the ``up-anti-up'' and ``down-anti-down'' quark pairs to be
greater than 70 GeV (where ``up'' stands for $u$ and $c$ and ``down'' for $d$
and $s$), so as to eliminate the soft photon contributions to pair production.
\bfig
\bce
\epsfig{file=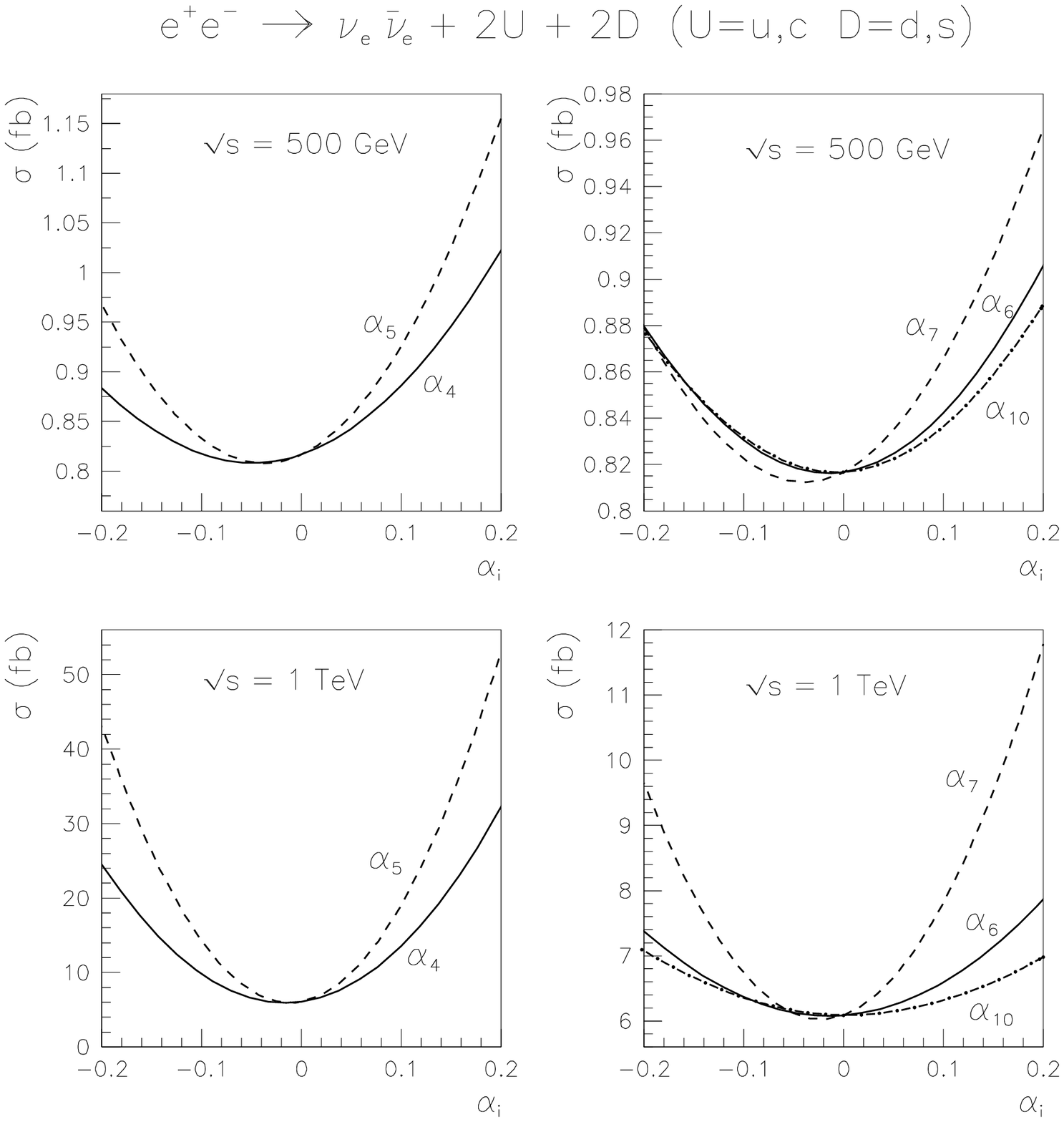,height=12.cm,width=12.cm}
\caption{\small Cross-section in the Born approximation as a function of
 the anomalous coupling parameters. The invariant masses of the pairs of
 up-anti-up and down-anti-down quarks are greater than 70 GeV.}
\label{fig:asud70}
\ece
\efig

In the plots of fig.~\ref{fig:asud70} the \xs\ is shown as a function of the
various parameters $\alpha_i$ at the two energies of 500 GeV and 1 TeV.
First of all, it should be observed that the effects of the anomalous couplings
are greater at 1 TeV than at 500 GeV, as expected, according to the above
arguments on unitarity violation. As a consequence, the \xs\ turns out to
have a very strong growth with energy for large values of the anomalous
coefficients. The growth of the \xs\ with energy is in agreement with results
for the scattering of real longitudinal gauge bosons obtained by means of the
equivalence theorem~\cite{qgcm1}. Moreover it can be observed that the greatest
effects are given by the coefficients $\alpha_4$ and $\alpha_5$, while the
coefficient $\alpha_{10}$ gives the smallest effect. To understand this fact it
is useful to examine the expressions of the operators in
eqs.~(\ref{eq:2.6}--\ref{eq:2.10}), where it can be seen that $\cl_4$ and
$\cl_5$ are the only terms where the vertex $4W$ appears. This vertex gives a
greater enhancement with respect to the others due to the fact that the
couplings of the $W$ boson to the fermions are stronger than those of the $Z$
boson. For the same reason, the operator $\cl_{10}$ has the smallest effect,
since it contributes only to the $4Z$ vertex. Moreover the $4Z$ vertex, as can
be seen in fig.~\ref{fig:agcd0}, can appear only in $s$-channel diagrams, and
has not the advantage of the $t$-channel growth at high energy.

The parameters $\alpha_6$, $\alpha_7$ and $\alpha_{10}$, that induce violation
of the custodial symmetry, are more constrained by the radiative corrections
to the $\rho$ parameter with respect to $\alpha_4$ and $\alpha_5$. For this
reason, the latter are more interesting and the remaining part of the study
will be restricted to them.

The behaviour of the \xs\ as a function of the c.m. energy is considered
in fig.~\ref{fig:a45sscanr}, where the energy range between 500 GeV and 1 TeV
is studied for the case without anomalous couplings and for two choices of the
anomalous parameters taken as examples, $\alpha_4=0.05$ and $\alpha_5=0.05$.
The invariant masses of all the possible quark pairs are required to be above
60 GeV. The solid curves include initial-state radiation~\cite{sf} (ISR) and
beamstrahlung~\cite{circe} (BS), while the dashed ones are in the Born
approximation. The growth with energy is already present when the parameters
$\alpha_i$ are equal to zero, and is enhanced when they are different from zero.
The effect of ISR and BS is to reduce the effective c.m. energy, and this
explains the observed lowering of $15-20 \%$ in the \xss\ with respect to the
Born approximation.
\bfig
\bce
\epsfig{file=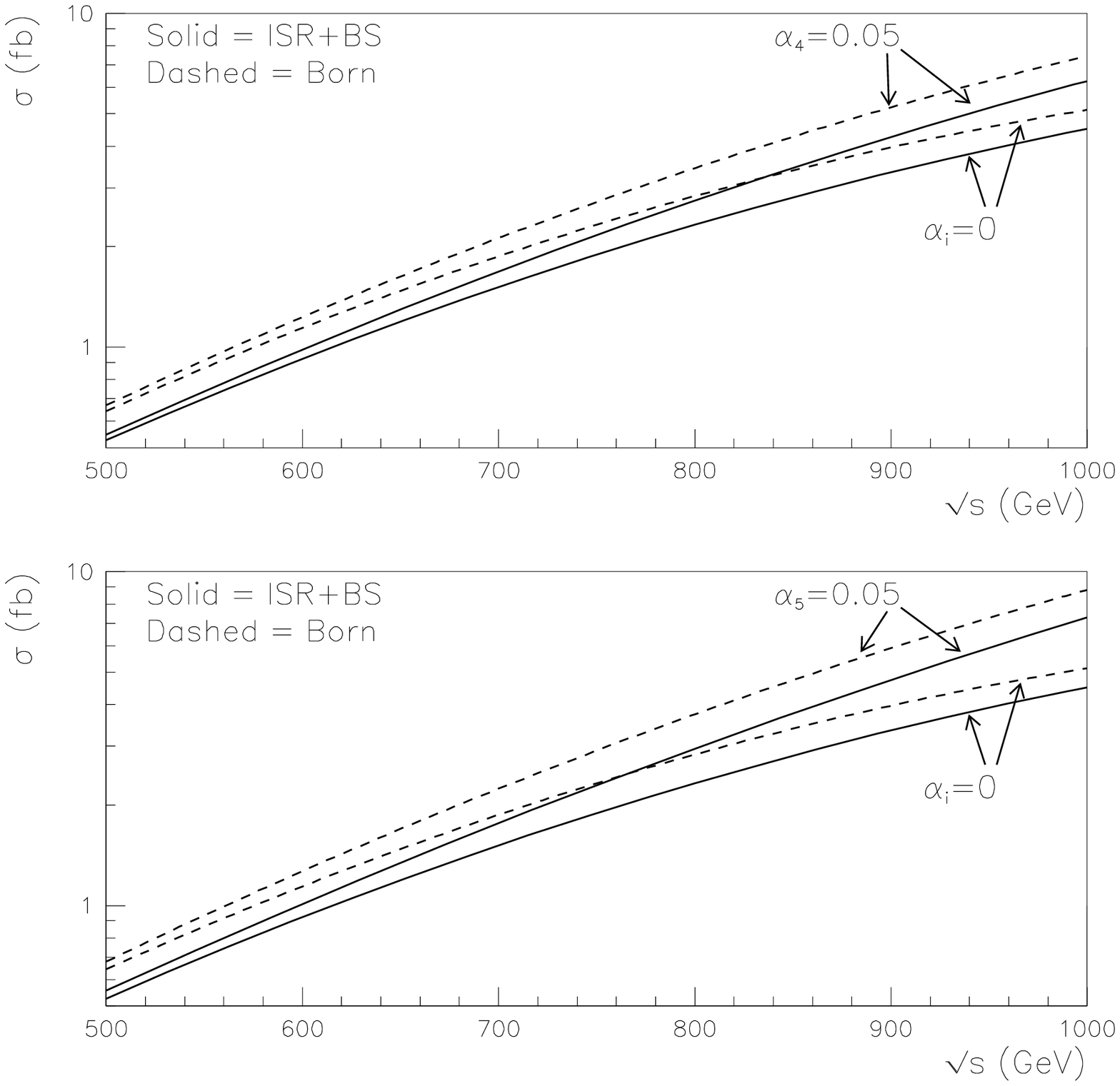,height=12.cm,width=12.cm}
\caption{\small Integrated \xs\ with initial-state radiation and beamstrahlung
 (solid lines) and in the Born approximation (dashed lines) as a function of
 the c.m. energy. The result with all the coefficients $\alpha_i$ set to zero is
 compared with the cases $\alpha_4=0.05$ (first plot) and $\alpha_5=0.05$
 (second plot).}
\label{fig:a45sscanr}
\ece
\efig

In order to study the sensitivity of some differential distributions to the
anomalous couplings, two samples of events have then been generated, one with
all the $\alpha_i$ equal to zero and the other with $\alpha_4=0.1$. The c.m.
energy is 1 TeV and the numbers of events correspond to a luminosity of 1000
fb$^{-1}$. As in fig.~\ref{fig:a45sscanr}, all the quark pairs have an
invariant mass greater than 60 GeV. A relatively large value of $\alpha_4$ has
been chosen, so as to have sizeable effects: indeed, the objective of this
analysis is to obtain clear indications on the cuts to be applied for enhancing
the sensitivity to the anomalous couplings under study. Among the variables
considered, the ones that turn out to be most sensitive to the parameter
$\alpha_4$ are defined in the following way. The invariant mass of the system
of four jets is indicated as $M(WW)$. This variable does not require any
identification procedure. The other variables that have been studied are based
instead on simple identification algorithms, to take into account the fact that
the quark flavours are not distinguishable. The $W$ bosons are reconstructed as
the pairs of quarks $q_iq_j$ and $q_kq_l$ such that the quantity
$|m(q_iq_j)^2-M_W^2|+|m(q_kq_l)^2-M_W^2|$ is minimized. The angle of one
reconstructed $W$ boson with respect to the beam axis is indicated as
$\theta(W)$. For each event, the four jets are ordered in transverse momentum
and are labelled with $j_1,\ldots,j_4$, where $j_1$ is the one with greatest
$p_t$. The invariant masses of pairs of jets are then considered, and one in
particular, $M(j_3j_4)$, that is the invariant mass of the pair of jets with
lowest $p_t$, is found to be most sensitive to the anomalous couplings under
study. In fig.~\ref{fig:a4ntr} the variables defined above are shown for the
sample with $\alpha_4=0.1$ (solid histograms) in comparison with the sample
where all the anomalous parameters are set to zero (dashed histograms). In
these plots, where the different total numbers of events for the two samples,
due to the differences of the \xss, must be taken into account, significant
deviations are found in the shapes of the distributions. This is due to the
Lorentz structures of the anomalous vertices involved, that tend to populate
the region of phase-space with high transverse momenta for the gauge bosons.
\bfig
\bce
\epsfig{file=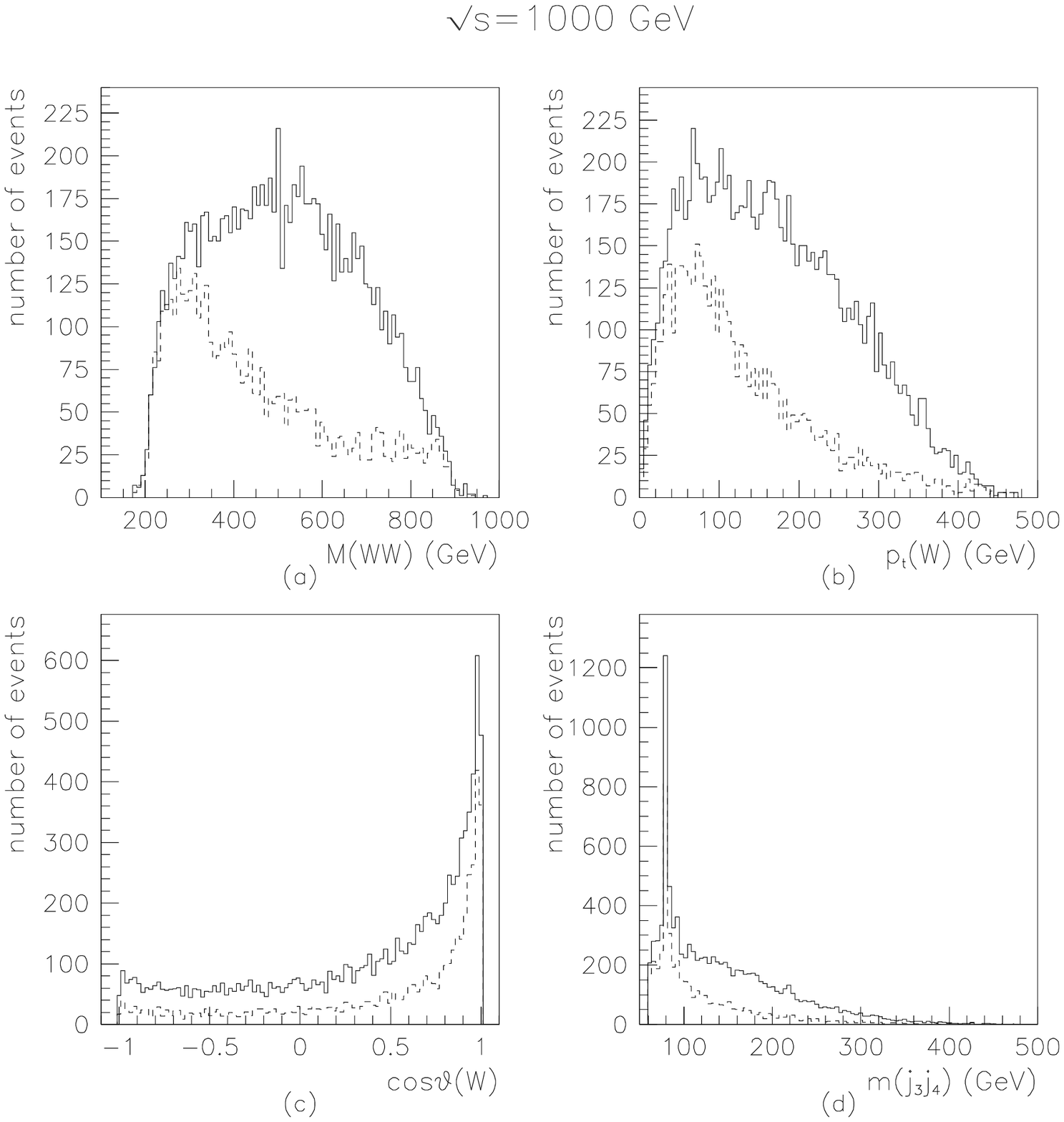,height=12.cm,width=12.cm}
\caption{\small Comparison between two samples of events with $\alpha_4=0.1$
 (solid histograms) and $\alpha_i=0$ (dashed histograms) in the presence of
 initial-state-radiation and beamstrahlung. The samples are
 normalized to a luminosity of 1000 fb$^{-1}$, and have different numbers of
 events, due to the different \xss\ . (a): invariant mass
 distribution of the system of four jets; (b): distribution of the transverse
 momentum of the reconstructed $W$ boson; (c): angular distribution of the
 reconstructed $W$ boson with respect to the beam axis; (d): invariant mass
 distribution of the pair of quarks with lowest transverse momenta.}
\label{fig:a4ntr}
\ece
\efig
In view of a realistic simulation, these distributions include the effects of
ISR and BS. It has been verified that these effects do not introduce significant
variations with respect to the Born approximation, as can be expected, since
the variables considered do not involve the momenta of the neutrinos.

On the contrary, the distribution of the missing mass, defined as
$M_{miss}=\sqrt{P_{miss}^2}$, where $P_{miss}$ is the missing momentum, is
strongly affected by ISR and BS. This variable, that in the Born approximation
is the invariant mass of the neutrino pair, is shown in fig.~\ref{fig:mmiss}:
in the upper plot the case with $\alpha_i=0$ is considered, while the lower plot
refers to $\alpha_4=0.1$. The solid histograms include ISR and BS, the dashed
histograms are in the Born approximation. In both plots the peaks corresponding
to the $Z$ mass are almost cancelled by the effects of ISR and BS.

\bfig
\bce
\epsfig{file=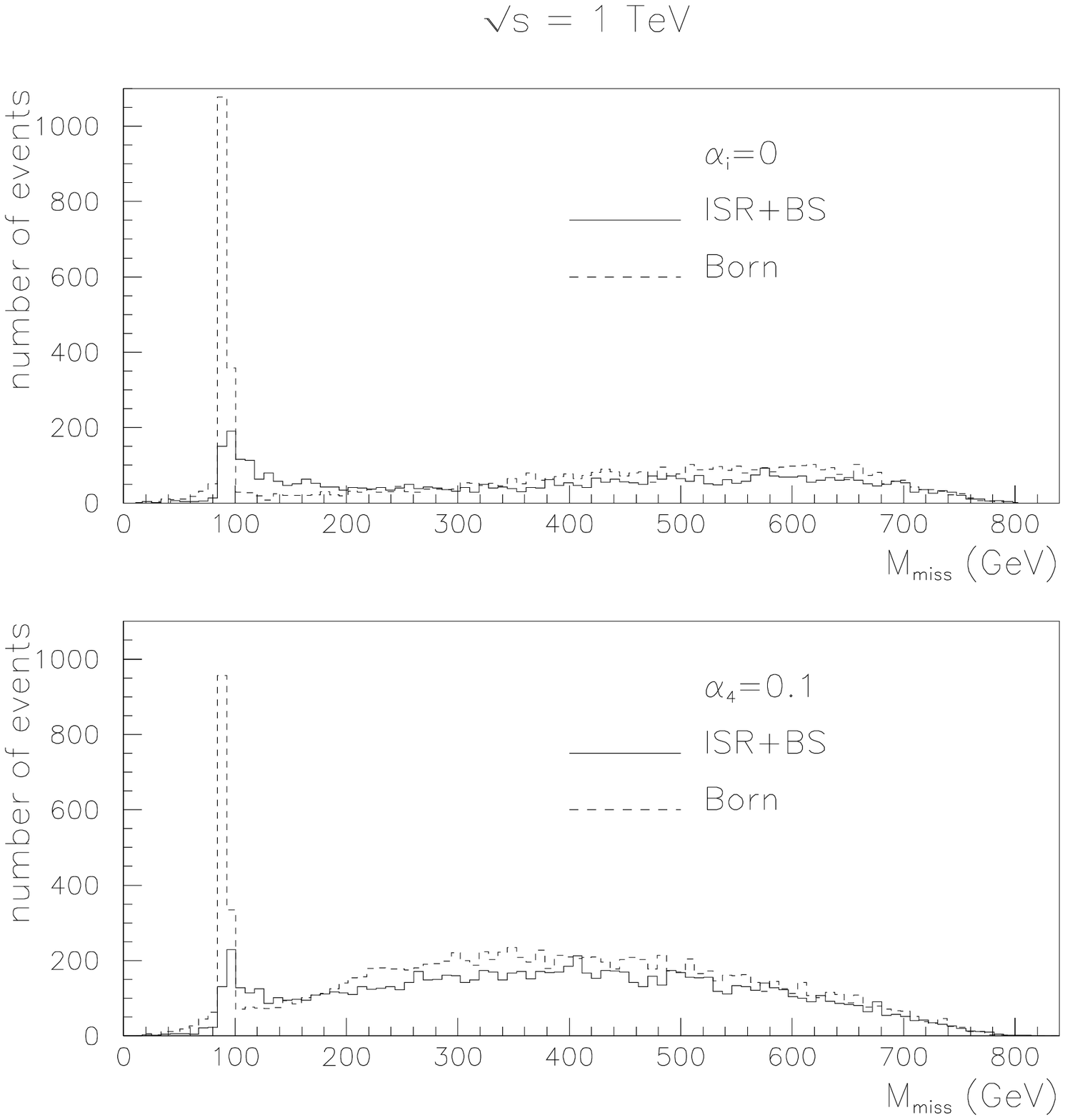,height=12.cm,width=12.cm}
\caption{\small Missing mass distribution. Upper plot: $\alpha_i=0$, lower plot:
 $\alpha_4=0.1$. Solid: ISR+BS; dashed: Born approximation.}
\label{fig:mmiss}
\ece
\efig

Results very similar to those illustrated above are obtained for the coupling
$\alpha_5$. On the basis of this analysis of distributions, suitable cuts have
then been determined to enhance the sensitivity to the anomalous couplings
$\alpha_4$ and $\alpha_5$. The results are given in fig.~\ref{fig:a45fine},
where the \xs\ in the Born approximation is shown as a function of the
parameters $\alpha_4$ (upper plot) and $\alpha_5$ (lower plot) in the range
$(-0.01,0.01)$.

\bfig
\bce
\epsfig{file=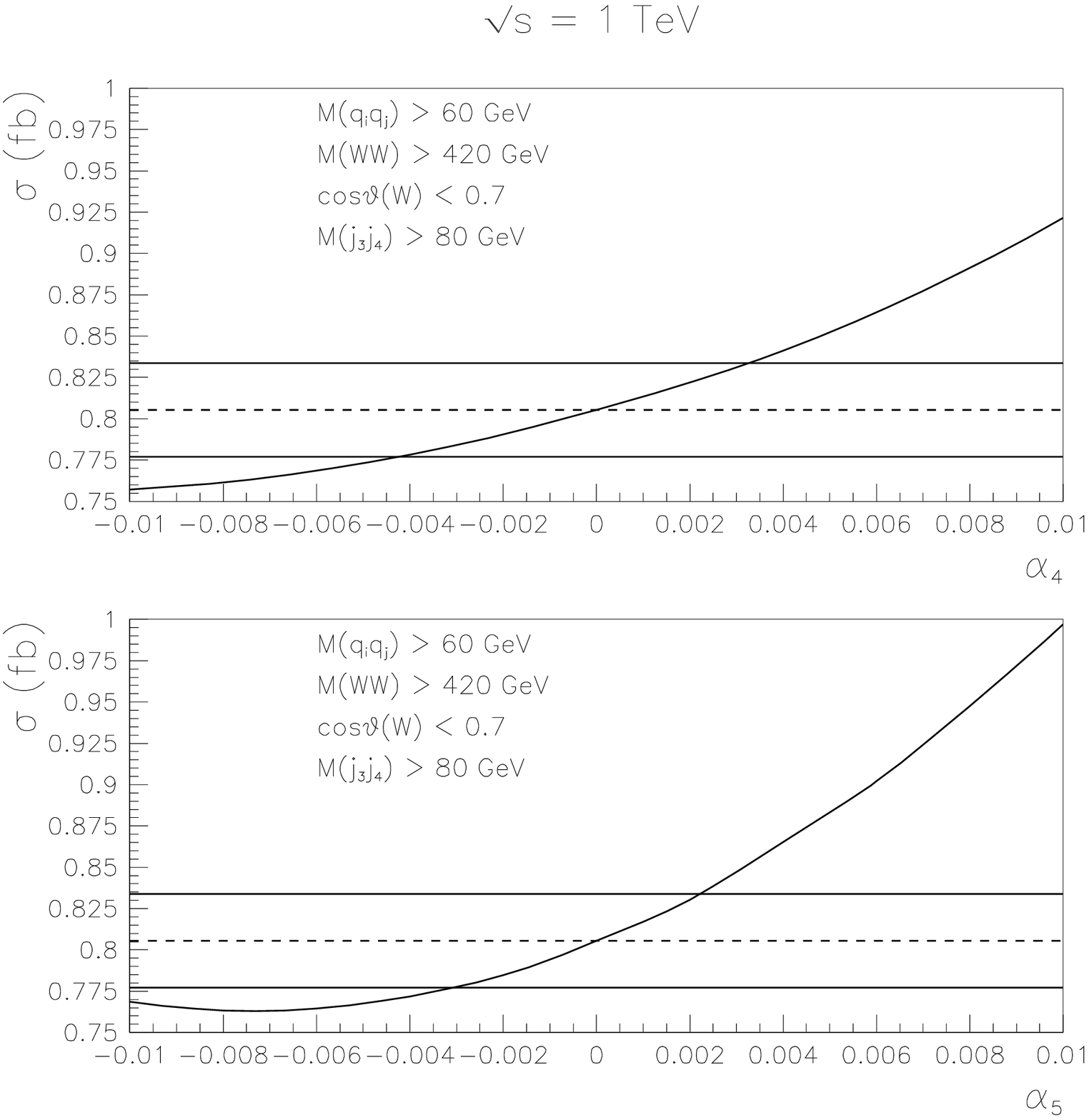,height=12.cm,width=12.cm}
\caption{\small Integrated \xs\ in the Born approximation at a c.m. energy of
 1 TeV, as a function of $\alpha_4$ (first plot) and
 $\alpha_5$ (second plot) with a set of kinematical cuts studied to
 enhance the sensitivity to the anomalous couplings. The horizontal solid lines
 are the $1 \sigma$ bounds around the $\alpha_i=0$ value (dashed line), with
 the experimental uncertainty corresponding to a luminosity of 1000 fb$^{-1}$.}
\label{fig:a45fine}
\ece
\efig
The cuts are defined as follows: $M(WW)$ is greater than 420 GeV,
the invariant mass of the two jets with the lowest transverse momenta,
$m(j_3j_4)$, is greater than 80 GeV and the angle of one reconstructed $W$
boson with respect to the beam axis satisfies $\cos\theta(W)<0.7$. Moreover,
the invariant mass of all the quark pairs is greater than 60 GeV. It has been
found that neither the addition of a cut on the variable $p_t(W)$ nor its
substitution to one of the above cuts can improve the results obtained.
In the same plots, the $1\sigma$ limits around the value of the \xs\ for
$\alpha_i=0$, with an experimental error corresponding to a luminosity of 1000
fb$^{-1}$, are shown. It can be seen that the sensitivity to the parameters
$\alpha_4$ and $\alpha_5$ turns out to be in the range between $10^{-3}$
and $10^{-2}$. These conclusions are not modified by the introduction of ISR
and BS, since the variables used in the cuts are not sensitive to these effects.
\section{Conclusions}
\label{sect:4}
The analysis of anomalous dimension-four quartic gauge couplings involves
processes with at least six fermions in the final state. A set of processes of
this kind has been considered in this work. By using a Monte Carlo event
generator that has already been employed for other phenomenological studies on
$6f$ physics, complete tree-level results have been obtained, in the
context of the electroweak chiral lagrangian formalism. The five dimension-four
operators giving genuine quartic anomalous couplings have been considered, and
a special attention has been devoted to the two custodial-symmetry conserving
ones, usually indicated by $\cl_4$ and $\cl_5$ and involving the parameters
$\alpha_4$ and $\alpha_5$. After studying the energy dependence of the \xs\ and
the effects of initial-state-radiation and beamstrahlung, a set of kinematical
cuts has been considered in order to enhance the signals of the anomalous
couplings. A sensitivity to the parameters $\alpha_4$ and $\alpha_5$ in the
range between $10^{-3}$ and $10^{-2}$ has been found at 1 TeV.

The results discussed above can be seen as an example of the possibilities of
study of these phenomena through a complete tree-level simulation of $6f$
processes. In order to make the calculations quantitatively more accurate, some
simplifications that have been adopted here should be eliminated: namely the
class of signatures to be taken into account should include a full sum over the
indistinguishable neutrino and quark flavour combinations in the final state,
and the effect of QCD backgrounds should be considered. In particular the latter
improvement can be achieved by means of the last version of ALPHA~\cite{alpha1},
that includes the QCD lagrangian. Moreover, the role of polarization of the
initial electrons and positrons could be investigated. These points will be the
the subject of future developments.

\vspace{1.truecm}
\noindent
{\bf Acknowledgements}\\
The author would like to thank G.~Montagna, M.~Moretti, O.~Nicrosini and
F.~Piccinini for stimulating discussions and useful comments on the manuscript.
The author also thanks the INFN, Sezione di Pavia, for the use of computing
facilities.

\end{document}